\documentclass{solarphysics}
\usepackage{graphicx}        
\usepackage[usenames]{color}    
\usepackage[optionalrh,solaenum]{spr-sola-addons} 
\usepackage[urlcolor=blue]{hyperref}
\ifx \doiurl    \undefined \def \doiurl#1{\href{http://dx.doi.org/#1}{\textsf{DOI}}}\fi
\ifx \adsurl    \undefined \def \adsurl#1{\href{http://adsabs.harvard.edu/abs/#1}{\textsf{ADS}}}\fi
\ifx \arxivurl  \undefined \def \arxivurl#1{\href{http://arxiv.org/abs/#1}{\textsf{arXiv}}}\fi

\newcommand{\aap}{    {\it Astron. Astrophys.}}

\newcommand{\aapr}{   {\it Astron. Astrophys. Rev.}}

\newcommand{\apj}{    {\it Astrophys. J.}}

\newcommand{\solphys}{{\it Solar Phys.}}

\newcommand{\figref}[1]{Figure \ref{#1}}

\renewcommand{\S}{Section }

\newcommand\arcsec{\mbox{$^{\prime\prime}$}}

\newcommand{\fnc}[1]{\textsf{#1}} 
\newcommand{\unit}[1]{\ensuremath{\, \mathrm{#1}}}

\begin{document}
\begin{article}
\begin{opening}
\runningauthor{L.A. Tarr \textit{et al.}}
\runningtitle{Quiescent Reconnection Rate}

\author{Lucas~A.~Tarr \sep Dana~W.~Longcope \sep David~E.~McKenzie \sep Keiji~Yoshimura}
\institute{Department of Physics, Montana State University, Bozeman, Montana 59717, USA\\
\href{mailto:ltarr@physics.montana.edu}{ltarr@physics.montana.edu}}
\title{Quiescent Reconnection Rate Between Emerging Active Regions and Preexisting Field, with Associated Heating: NOAA AR11112}

\begin{abstract}
  When magnetic flux emerges from beneath the photosphere it displaces the preexisting field in the corona, and a current sheet generally forms at the boundary between the old and new magnetic domains.  Reconnection in the current sheet relaxes this highly stressed configuration to a lower energy state.  This scenario is most familiar, and most often studied, in flares, where the flux transfer is rapid.  We present here a study of steady, quiescent flux transfer occurring at a rate three orders of magnitude below that in a large flare.  In particular we quantify the reconnection rate, and related energy release, occurring as new polarity emerges to form Active Region 11112 (\textsf{SOL16 October 2010T00:00:00L205C117}) within a region of preexisting flux.  A bright, low lying kernel of coronal loops above the emerging polarity, observed with the \emph{Atmospheric Imaging Assembly} onboard the \emph{Solar Dynamics Observatory} and the \emph{X-ray Telescope} onboard \emph{Hinode}, originally shows magnetic connectivity only between regions of newly emerged flux when overlaid on magnetograms from the \emph{Helioseisimic and Magnetic Imager}.  Over the course of several days, this bright kernel advances into the preexisting flux. The advancement of an easily visible boundary into the old flux regions allows measurement of the rate of reconnection between old and new magnetic domains.  We compare the reconnection rate to the inferred heating of the coronal plasma. To our knowledge, this is the first measurement of steady, quiescent heating related to reconnection.  We determine that the newly emerged flux reconnects at a fairly steady rate of $0.38 \times 10^{16}\unit{Mx \ s^{-1}}$ over two days, while the radiated power varies between $(2\sim8)\times 10^{25}\unit{erg \ s^{-1}}$ over the same time.  We find that as much as $40\%$ of the total emerged flux at any given time may have reconnected.  The total amount of transferred flux ($\sim1\times10^{21}\unit{Mx}$) and radiated energy ($\sim7.2\times10^{30}\unit{ergs}$) are comparable to that of a large M- or small X-class flare, but are stretched out over 45 hours.
\end{abstract}
\end{opening}
\section{\label{sec:intro4}Introduction} 

When magnetic flux emerges through the photospheric boundary, the coronal field must respond in some way.  The exact type of response will depend on the configuration of the preexisting field as well as the rate and total amount of flux emergence.  The response itself may be broken down into several stages \cite{Heyvaerts:1977}: a preflare heating phase, an impulsive phase of particle acceleration and rapid increase in H$\alpha$ emission, and finally a main phase where such emission decreases.  Such a model naturally accounts for many diverse observations, from quiet-Sun X-ray bright points to the largest observed flares with associated coronal mass ejections (CMEs).  

Much of the work of the last several decades has focused on the impulsive and decay phases.  This is partly because of their extravagant nature and direct impact on space weather at Earth, but also, and importantly, because of their relatively short timescale, typically less than one hour, and corresponding ease of observation.  In contrast, the preflare heating phase of flux emergence, a period marked by continuous magnetic reconnection between the new and old flux, may last for days.  An accurate description of the process requires simultaneous observations of the magnetic field, to capture the emergence itself, and of EUV and X-ray emission, to capture the coronal response.  Only in the last few years have such simultaneous and continuous observations at the needed spatial resolution been possible.

Magnetic reconnection is one of the most likely direct sources for coronal heating \cite{Archontis:2008,Reale:2010}.  In terms of the coronal energy balance, reconnection converts free magnetic energy -- energy in excess of the potential field defined by photospheric sources of flux -- into kinetic and thermal energy of the plasma.  It is easy to demonstrate that free magnetic energy rapidly increases during flux emergence if no reconnection occurs: the coronal field develops a tangential discontinuity, or current sheet, at the interface between the preexisting and newly emerged flux systems.  Field lines on one side of the discontinuity have footpoints wholly within the preexisting flux system and field lines on the other side wholly within the newly emerged system.  At the location of reconnection, field lines from the two sides exchange footpoints, so that two new field lines are created, each connecting new to old flux.  These new field lines retract, adding kinetic energy, and compress, adding thermal energy to the plasma \cite{Reeves:2008,Guidoni:2010}.

The properties of the current sheet are determined by the amount of emerging flux and its configuration relative to the preexisting coronal field \cite{Heyvaerts:1977,Archontis:2008}.  The current sheet itself determines the rate of magnetic reconnection and therefore the rate at which magnetic free energy is converted into kinetic and thermal energy.  In this way, the observed heating and reconnection rates of an emerging flux tube will help us understand the coronal response to emerging flux.

During a flare, these processes are driven by rapid reconnection, resulting in the observational feature of flare ribbons, typically observed in either H$\alpha$ or 1600\,\AA{} data \cite{Fletcher:2001,Qiu:2009}.  Such studies of rapid reconnection during flares, in particular those using the evolution of flare ribbons overlaid on cotemporal magnetograms \cite{Forbes:1984b,Poletto:1986,Fletcher:2001,Qiu:2002,Qiu:2007,Longcope:2007,Qiu:2009,Longcope:2010,Kazachenko:2010,Kazachenko:2012}, are the most direct antecedents of the present study.  These two-ribbon flares generally follow the standard CSHKP flare model (See \opencite{Priest:2002} for a review).  Reconnected field lines within the current sheet link directly to photospheric footpoints, so that bright flare ribbons outline magnetic flux whose field lines attach to the reconnection site.  The expansion of the ribbons therefore provides an estimate of both the total flux involved in an event and the rate at which it is transferred across the current sheet.  In a flare, the flux transfer and resulting energy release typically take about $30-60\unit{min}$.

If the energy released by reconnection is insufficient to substantially increase chromospheric UV emission above its background state then we cannot use flare ribbons to determine the rate of reconnection.  We refer to this situation as non-flaring reconnection, reconnection typified by a lack of UV flare ribbons, cotemporal integrated GOES X-ray emission, or rapidly brightening coronal loops.  A number of observational studies have estimated reconnection that is not directly associated with flares, with most inferring such reconnection by indirect means.  \inlinecite{Schrijver:1997} introduced the concept of the ``magnetic carpet'' to describe the evolution of magnetic flux in quiet sun regions.  New flux constantly emerges, fragments, merges, and cancels within the quiet sun, but the total amount of flux remains essentially constant.  This behavior implies a substantial amount of continual reconnection as has been examined by \inlinecite{Hagenaar:2001,Hagenaar:2003,Close:2005,Hagenaar:2008}.  Current estimates place the rate of recycling at $1-2\unit{hrs}$, although \inlinecite{Hagenaar:2008} show that the cadence of observation strongly affects the estimated rate.

\inlinecite{Longcope:2005b} provided a more direct measurement of non-flaring reconnection by studying the amount of reconnection between two separately emerged active regions.  As those authors note, coronal loops connecting two active regions have long been held as compelling evidence for reconnection within the corona.  They therefore used EUV observations of coronal loops between a freshly emerged active region and a nearby preexisting active region to quantify reconnection between the two regions.  By counting loops, the authors were able to estimate the flux contained in a single coronal domain of interest.  In a potential field model, this domain contained a certain finite amount of flux, while the observational history of the active region complex, namely the emergence of one entire active region, suggests that, at least initially, that domain contained identically zero flux in the actual coronal field.  Any loops observed in that coronal domain then demonstrate that reconnection in the corona has transferred flux into that domain.  One necessary drawback of the study was that it used some simplifying assumptions about the amount of flux represented by each observed coronal loop.  Essentially, each ``individual coronal loop'' was assumed to represent a unit amount of transferred flux, about $\sim 4\times 10^{18}\unit{Mx}$, rather than following the loop to its footpoint and determining the underlying amount of flux from a coaligned magnetogram.  

We apply similar methodologies to those mentioned above to high resolution data from the \emph{Atmospheric Imaging Assembly} (AIA: \opencite{Lemen:2012}) and \emph{Helioseismic and Magnetic Imager} (HMI: \opencite{Scherrer:2012}) instruments onboard the \emph{Solar Dynamics Observatory} spacecraft (SDO: \opencite{Pesnell:2012}).  We focus on NOAA active region 11112, a small region whose complete emergence occurs while on the Earthward solar hemisphere.  The polarity composing AR11112 itself emerges within a large, diffuse, preexisting unnamed active region.  \figref{fig:magcontext} provides a context image, with the top panel showing the radial magnetic field derived from HMI data prior to emergence, and the bottom panel showing a cospatial and cotemporal image from the AIA 211\,\AA{} channel.  For simplicity, we will refer to this entire region as AR11112 and explicitly distinguish between the emerged and preexisting field.

As the new polarity emerges, a bright ellipsoidal kernel of loops forms, interconnecting the emerged flux.  The kernel is visible in various EUV channels of AIA data, as well as in data from the \emph{X-ray Telescope} onboard the \emph{Hinode} satellite \cite{Kosugi:2007}.  The magnetograms and coronal imagery allow us to study the interplay of the emerging magnetic flux and surrounding coronal field.  In particular, the observed coronal response in terms of magnetic connectivity and radiated power shed light on the energetic consequences of the emergence.  We focus on the slow, quiescent response prior to a GOES M2.9 class flare on 16 October 2010 19:00UT.  Interested readers may refer to \inlinecite{Tarr:2012} for an analysis of magnetic field topology and energetics pertaining to the flare itself.

In Section \ref{sec:data} we describe the observations.  Section \ref{sec:tracking} explains our use of the magnetic field data to characterize the emerging flux region.  Section \ref{sec:euv} details how we use the EUV observations to determine the actual coronal flux domains.  In Section \ref{sec:rxrate} we combine the EUV and magnetic field observations to find the amount of reconnected flux at every time in our data series.  In Section \ref{sec:xrt} we describe the use of XRT filter ratios to calculate the radiated power within the region of interest.  Finally, in Section \ref{sec:discussion} we discuss the results of our observations and how this informs our model of the flux emergence process.

\section{\label{sec:data}Data}

AR11112 begins a small episode of magnetic flux emergence at 14 October 2010 05:00\footnote{Here and throughout, all times are stated in UT.}, centered on (13East, 20South).  The episode lasts for several days, culminating in a GOES class M2.9 flare at 16 October 2010 19:00.  The emergence takes place completely within a region of preexisting negative flux, as indicated in \figref{fig:magcontext} top.  This figure shows the radial magnetic field for the entire diffuse active region in plane-of-sky (POS) coordinates 5 hours before emergence.  The radial field is derived from measurements of the photospheric vector magnetic field from the SDO/HMI instrument, obtained via the \textsf{hmi.sharp\_720s} series at JSOC, SHARP number 0211.  HMI has $0.5\arcsec \times 0.5\arcsec$ pixels.  The $180^\circ$ ambiguity in the azimuth has been disambiguated using a variant of the Metcalf minimization scheme as described in \inlinecite{Sun:2012}.  

The magnetic field data series last for 71 hours, from 13 October 2010 23:58 to 16 October 2010 22:58, at a cadence of 1 measurement per hour.  Each pixel is transformed from POS to radial, longitude, latitude coordinates so that we may calculate the radial (solar surface normal) magnetic flux through each pixel.  Throughout this work, the greyscale in figures of magnetograms, such as \figref{fig:magcontext} top, is scaled to represent the radial flux through each pixel, in maxwells, while the image coordinates are kept in the POS system.  This facilitates easy comparison between the magnetic data and the EUV/X-ray data.

We use extreme ultraviolet (EUV) data from SDO/AIA to determine how the emerging polarity interacts with surrounding field.  We focus on data from the SDO/AIA 211\,\AA{} channel, as seen in the bottom panel of \figref{fig:magcontext}.  The AIA data are prepared to level 1.5 using the standard {\sf aia\_prep.pro} utility in SolarSoftware (SSW: \opencite{Freeland:1998}) and cover the same period of time as the HMI data.  Each 211\,\AA{} image is taken roughly 2 minutes after the closest--in--time HMI magnetogram, and the two data series are coaligned as discussed below.  The AIA instrument has $0.6\arcsec\times0.6\arcsec$ pixels.

Finally, we use two filters from \emph{Hinode}/XRT: the titanium-on-polyimide and thin-aluminum-on-mesh filters, colloquially referred to as `Ti--poly' and `Al--mesh.'  Because we use a ratio of the two filtergrams at each time to form temperature and emission measure maps, both filters must be present.  Our analysis is therefore limited to the Al--mesh filter's normal cadence of 1 image per hour over the course of these observations.  There are occasional multi--hour lapses as XRT either performs synoptic observations or points towards other features on the solar disk.  The XRT data is prepared using the \fnc{xrt\_prep.pro} utility in SSW, described in \inlinecite{Kobelski:2013}.

\subsection{\label{sec:ca}Co--alignment}
Spatially, coalignment between the SDO/AIA and XRT images was accomplished by calibrating the differences in roll angle, plate scale, and pointing between the two instruments.  The differences in the roll angles and plate scales were measured accurately through a cross calibration technique using full disk solar images (details are in Yoshimura and McKenzie (in preparation)).  The pointing differences were corrected by applying a cross correlation technique using AIA 335\AA{} and XRT thin-filter images.  Since all the instruments onboard SDO were well calibrated against each other by the Venus-transit observation in June 2012, we can coalign the XRT data with any data from AIA and HMI.  The error of the coalignment in this study is estimated to be smaller than 1 arcsec.

Our analysis uses approximately a one hour cadence.  If there is a time difference between two observations we translate all relevant images to the time of the nearest magnetic field data using the differential solar rotation rate described in \inlinecite{Snodgrass:1990}.

\section{\label{sec:tracking}Tracking Photospheric Flux Concentrations}
\begin{figure}
  \centerline{\includegraphics[width=\textwidth,clip=]{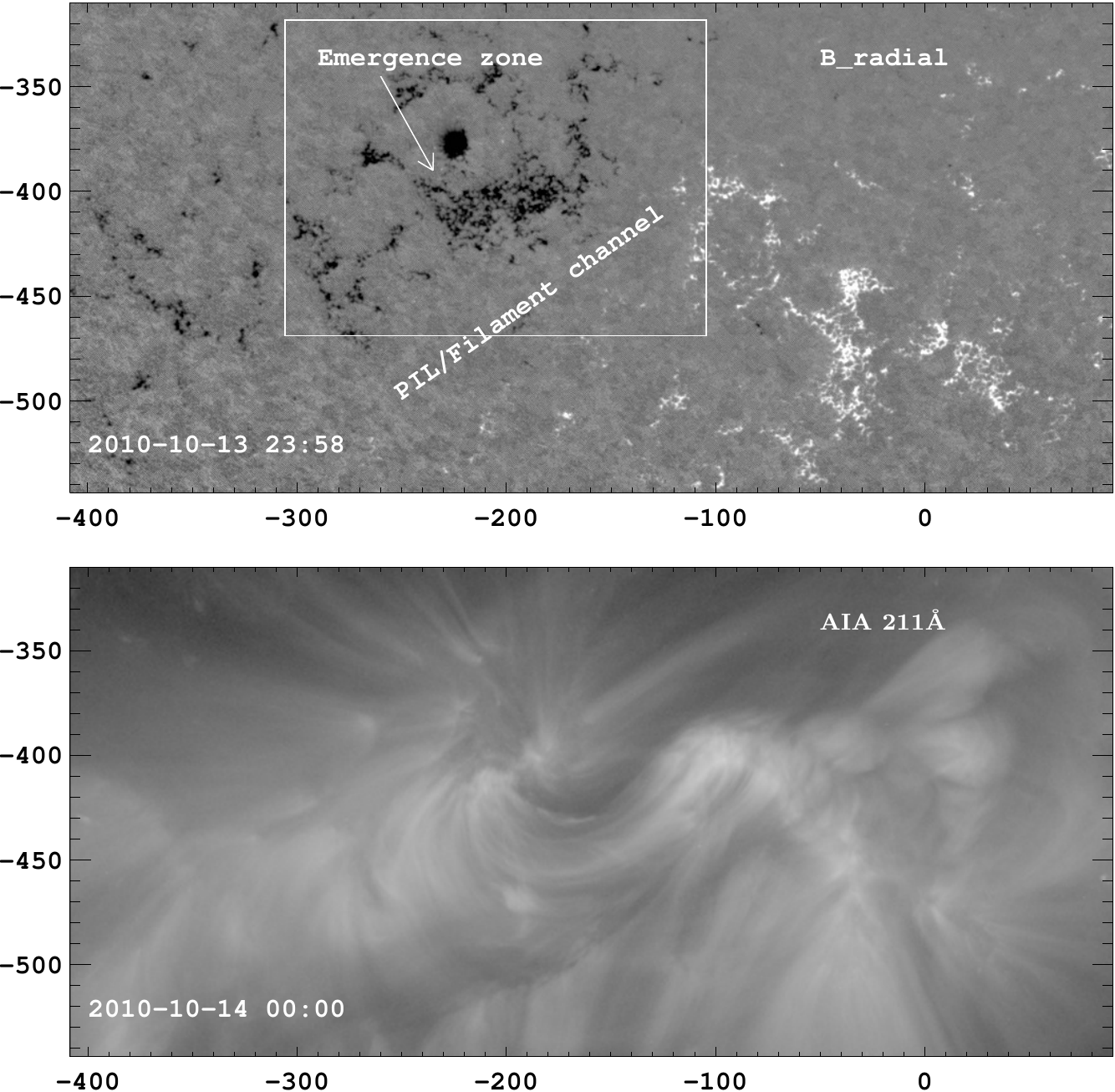}}
\caption[AR11112 radial magnetic field context]{Top: Context magnetogram showing radial magnetic flux through each pixel, displayed in POS coordinates, with x- and y-axes in arcseconds from disk center.  The greyscale saturates at $\pm 6.66\times 10^{17}\unit{Mx}$ (an average field strength of $500\unit{G}$ for a pixel located at disk center).  Light(dark) pixels are positive(negative) polarity.  The rectangle shows the field of view used throughout the analysis.  The large scale PIL bisects the area, and the emergence we discuss occurs within the diffuse negative polarity regions, centered slightly South and East of the ``bull's eye'' feature.  Bottom: Near-simultaneous log-scaled AIA 211\,\AA{} image of the same field of view.}
\label{fig:magcontext}
\end{figure}
The top panel in \figref{fig:magcontext} shows a large field of view (FOV) magnetogram of the preexisting diffuse active region.  A large polarity inversion line (PIL) separates the negative flux to the east from positive flux to the west.  EUV images from this time, such as the AIA 211\AA{} channel in the bottom panel of the figure, show a simple arcade of field lines arching over the PIL.  Visible in various AIA channels is a large filament laying along the PIL, apparently underneath the arcade.  The boxed region shows the FOV we will use for the remainder of our analysis.  All emergence occurs within this unipolar, negative polarity region.

We repeat the flux-tracking analysis of \inlinecite{Tarr:2012}.  That analysis used the 45-second cadence line-of-sight (LOS) magnetograms from HMI available at the time.  We now use the actual radial field derived from the vector magnetograms.  Our former study was a detailed analysis of the emerging field's topology and connectivity and the resulting energetics within the framework of the Minimum Current Corona model (MCC: \opencite{Longcope:2001}).  In contrast, in the present investigation we use the magnetic field to distinguish between all new and old flux of each polarity, which we can do with ease, but do not need in as precise detail as the for the previous study.  We therefore use a simplified mask array, and correspondingly different region labels.  

\begin{figure}[ht]
  \begin{center}
    \includegraphics[width=0.8\textwidth]{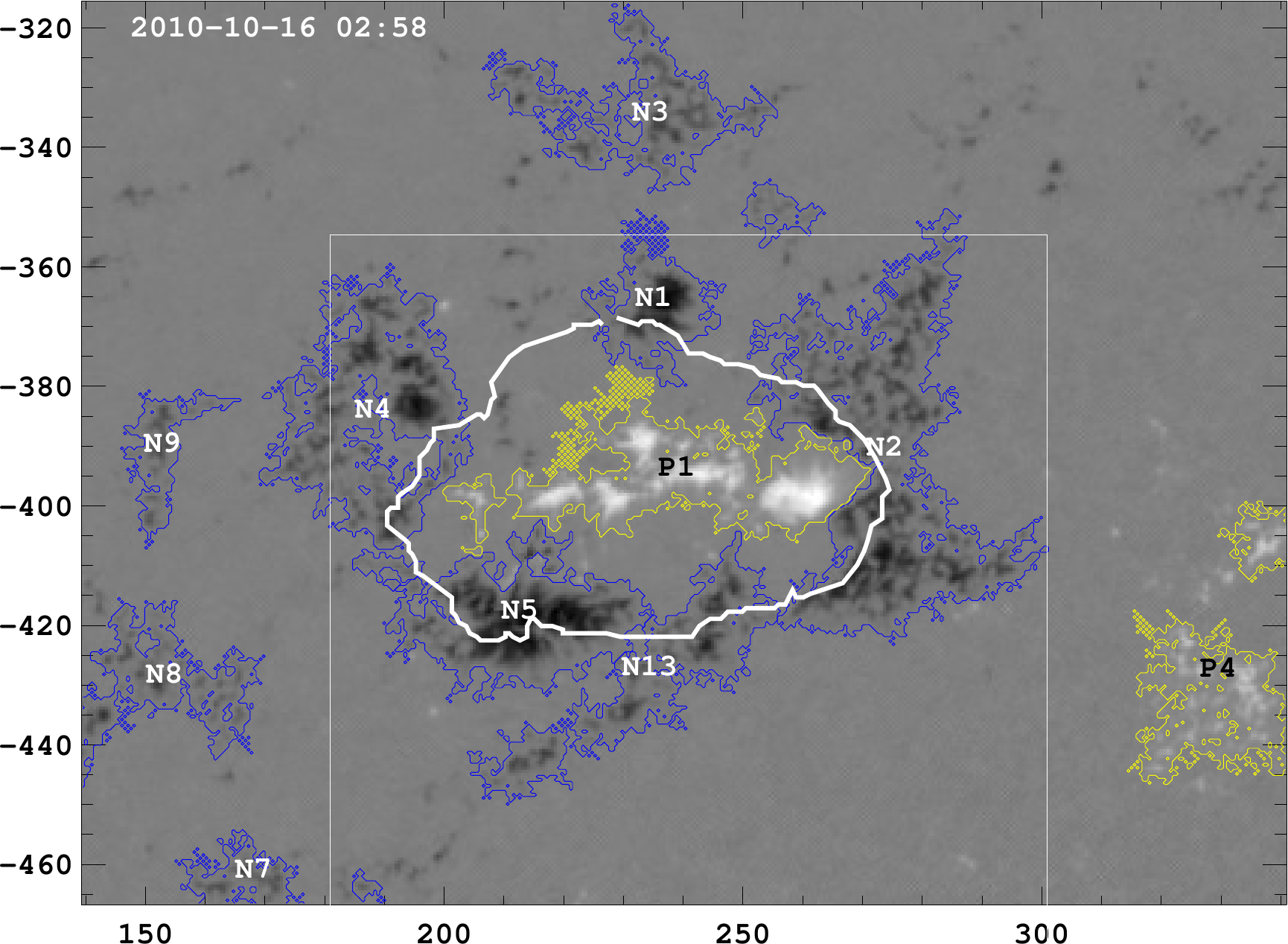}
    \caption[AR11112 radial magnetic field detail]{Same as \figref{fig:magcontext}, for the smaller FOV.  This greyscale saturates at $\pm 2.66\times 10^{18}\unit{Mx}$ (an average field strength of $2000\unit{G}$ for a pixel located at disk center).  Thin colored lines show the regions tracked in our mask array.  The thick line is the boundary derived from AIA 211\AA{} images, shown in \figref{fig:euvbnd}.  Region labels are marked at the flux-weighted centroid of each region.  Note that the large region of preexisting negative flux to the west of the emergence region, N2, now forms a crescent shape and its label therefore lies on the boundary between P1 and N2.  The box shows the FOV of \figref{fig:channelcomp}.}
  \label{fig:magevo}
  \end{center}
\end{figure}

With our slightly different analysis goal in the present work, we use slightly different algorithm parameters compared to \inlinecite{Tarr:2012}.  For this study, we use a lower threshold of $50\unit{G}$ average field strength over each pixel, and create a smoother mask structure using a potential field extrapolation from the lower (photospheric) boundary to a height $h=3.0\unit{Mm}$.  Each mask region must contain at least $2.66\times 10^{19}\unit{Mx}$ ($2\times 10^4 \unit{G*px}$), and exist for at least 5 hours, for inclusion in our analysis.  For comparison, near the end of our time series (Oct 16, 23:00) the total unsigned flux (above 50 gauss) for the large FOV magnetogram (cf. \figref{fig:magcontext}) is $3.04\times10^{22}\unit{Mx}$, while the total unsigned flux of newly emerged field is $4.09\times 10^{21}\unit{Mx}$, about $13\%$ of the whole.

The primary regions we discuss in this study are described as follows (see \figref{fig:magevo}).  In the region of emergence, we label all emerging positive flux as P1, rather than distinguishing between the different concentrations in successive episodes of emergence.  N1 is the circular ``Bull's Eye'' concentration of strong negative flux just north of emergence.  N4 and N5 are regions to the East and South of emergence, respectively, that contain difficult-to-distinguish mixes of new and old flux, although by the end of emergence both contain primarily new flux.  N13 is the oblong region just west of N5.  It contains primarily preexisting flux, but is located near the center of emergence, so we have grouped it separately.  

We will focus the most attention on the region labeled N2, just west of the emergence zone.  Prior to emergence, this region is roughly circular, consisting of numerous small flux concentrations.  The emerging field drives into the middle of this patch of plage, pushing parts of N2 to the North and South and deforming it into the crescent shape seen in \figref{fig:magevo} (cf. the boxed portion of \figref{fig:magcontext}).  The total flux of N2 remains relatively constant at $2\times 10^{21}\unit{Mx}$, varying above and below this value by about $0.14\times 10^{21}\unit{Mx}$, or $7\%$ of the average value.  This variation is mostly due to patches of network flux joining to and detaching from the western boundary of the N2 mask region, and is 3 orders of magnitude less than the amount of emerged flux.  We are therefore very confident that the mask region N2 contains only flux that predates emergence, and that any coronal loops we observe connecting N2 to the emerging positive flux P1 are due solely to reconnection between the two flux domains.

\begin{figure}[ht]
  \begin{center}
    \includegraphics[width=\textwidth]{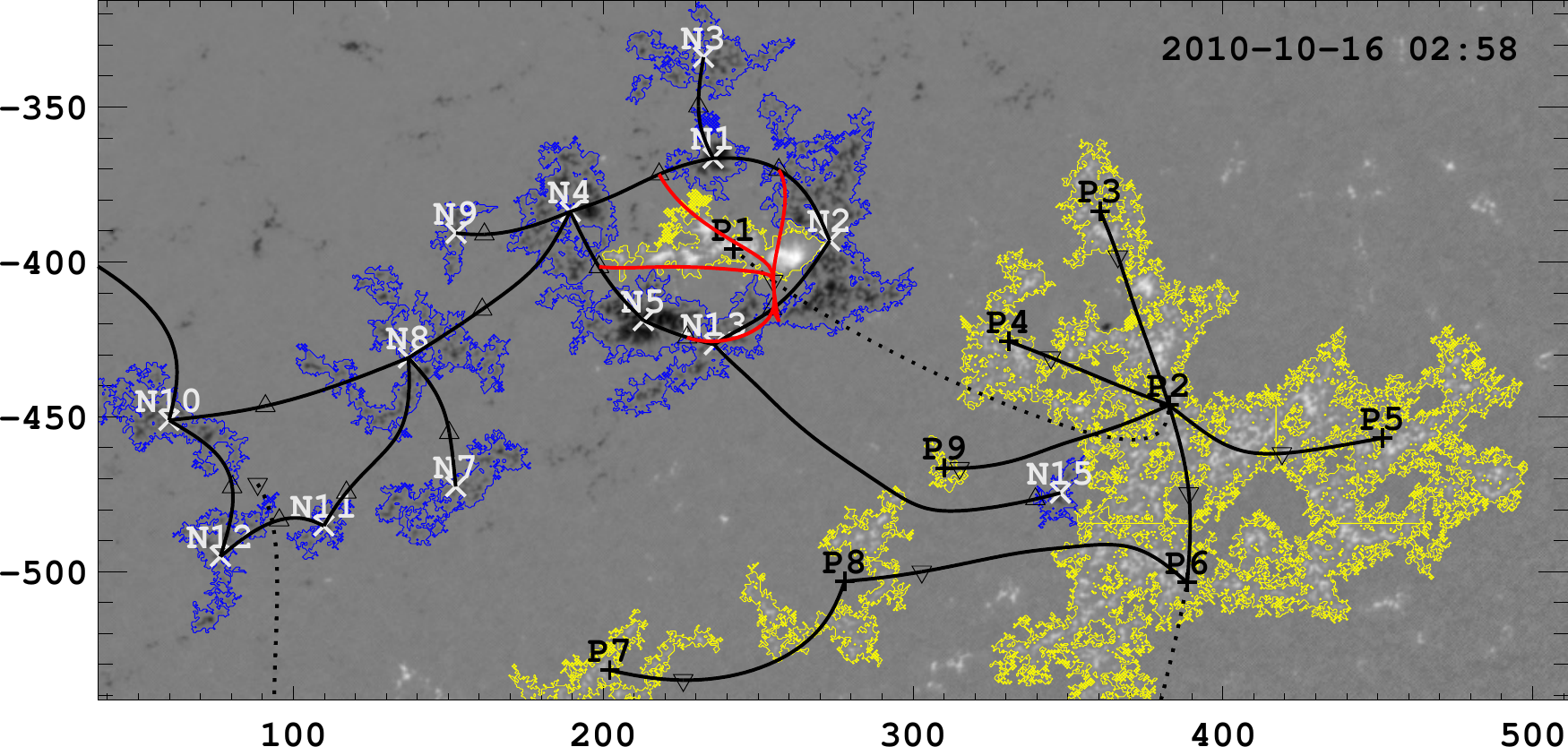}
    \caption[AR11112 radial field and topology]{Topology in an MCT model derived from the radial magnetic field, shown as the background image.  Magnetic charges are located at the flux weighted centroids of the mask regions, depicted as pluses (crosses) for positive (negative) regions.  Upright (inverted) triangles show A-type (B-type) nullpoints.  Thick black lines are spine field lines lying in the photospheric plane, while the dotted line is the spine line connected to the coronal null point located above the emerged positive polarity P1, near (260,-410).  The coronal null has one spine source in the emerged flux (P1), and one across the large scale PIL, in preexisting flux (P2).  Thick red lines are separators in the fan surface of the coronal null, connecting that null to the ring of spine field lines surrounding P1.  The radial field greyscale saturates at $\pm 2.66\times 10^{18}\unit{Mx}$.}
  \end{center}
\label{fig:topo}
\end{figure}

The topology of the region is described in detail in \inlinecite{Tarr:2012}.  Those authors use a Magnetic Charge Topology (MCT) to describe the field at each time, together with a MCC constraint to model time evolution.  An example MCT for the present work is shown in \figref{fig:topo}, with the various topological elements identified in the caption.  The essential element is that the ring of negative polarity field (regions N1, N2, N4, N5, and N13)\footnote{Note that the region labels are different in the present simplified study compared to \inlinecite{Tarr:2012}.} creates a ring of spine field lines connecting each source to nullpoints located between the sources.  This ring surrounds positive region P1.  P1 is one of the spine sources for a coronal null point, whose other spine source is located in the diffuse polar region to the west, in P2.  The fan (or separatrix surface) of the coronal nullpoint intersects the photosphere along the spine lines of the ring of negative polarity field, forming a dome over P1.  Field lines inside the dome connect to the newly emerged P1, while field lines outside the dome connect to preexisting positive field.  This is a commonly found topology in the solar corona, and as such has been studied numerically and theoretically in a number of ways (See \opencite{Pontin:2013} and references therein).

The animation of \figref{fig:magevo} shows the evolution of the radial magnetic field and our mask array.  While we transform the vector magnetic field data to find the radial, latitudinal, and longitudinal components at every pixel, throughout this study we display images in POS coordinates.  That is to say, in the magnetograms, the lightness or darkness of a pixel represents the value of total flux in $\unit{Mx}$ within the pixel, but we have not distorted the pixel's shape from CCD coordinates (this just makes the comparison of features observed in the HMI, AIA, and XRT data much more straightforward).

\section{\label{sec:euv}Magnetic Domains Observed in EUV}
\begin{figure}[ht]
  \begin{center}
    \includegraphics[width=\textwidth]{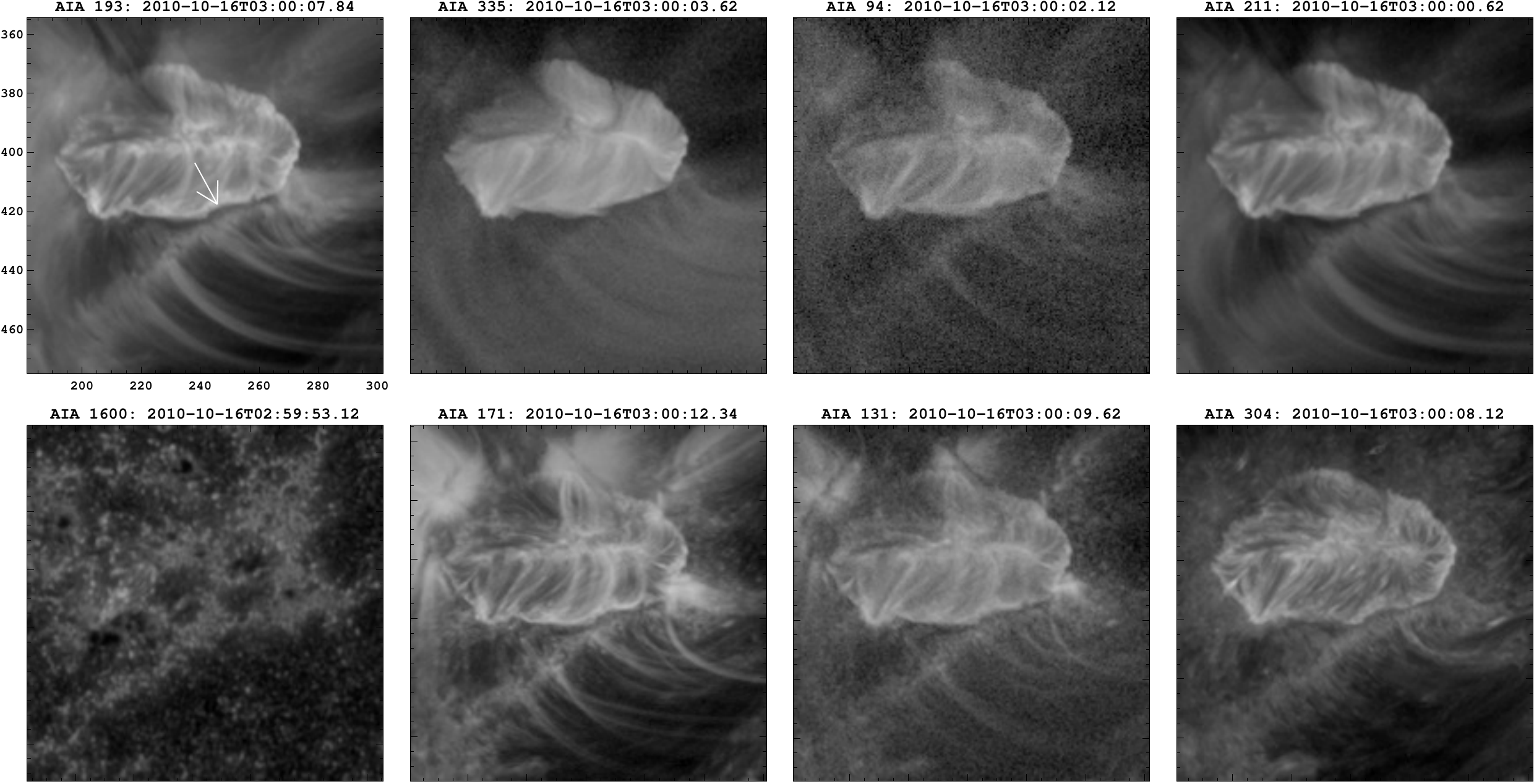}
    \caption[AR11112 detail of EUV channels]{Log-scaled UV and EUV data from AIA, for the timestep shown in \figref{fig:magevo} and FOV shown as the rectangle in that figure.  The dark band dividing the two flux domains, indicated by an arrow in the 193\AA{} (upper left) image, is visible in every EUV band, and from \figref{fig:magevo} we see that it surrounds the positive emerged flux, and lies completely within the negative flux regions.}
  \end{center}
\label{fig:channelcomp}
\end{figure}

As magnetic flux emerges from beneath the photosphere, we observe new magnetic loops in the EUV and X-ray images taken with AIA and XRT.  These appear as a bright cluster of short loops connecting the newly emerged positive and negative flux, as seen in \figref{fig:channelcomp} and the animation thereof.  The plasma that rises together with the magnetic field displaces the previously extant field in the coronal volume immediately above the photospheric emergence.  The displacement takes two forms.  The first is through horizontal shearing of the photospheric plasma to which the coronal field is anchored; the second is through generation of electric currents on the boundary between the old and new magnetic domains, which prevents the two from reconnecting.

The new flux emerges completely within an area of preexisting negative unipolar flux, with the PIL located to the west.  As discussed in \S\ref{sec:tracking}, this means the positive emerged flux is surrounded by a sea of negative flux, generating a magnetic dome topology, defined by the separatrix of a coronal nullpoint.  As illustrated in \figref{fig:topo}, one of the coronal null's spine sources is located underneath this dome (P1), while the other sits in the unipolar region of positive flux across the diffuse PIL (P2).  Field lines within the dome connect to newly emerged positive flux, while field lines outside it connect to preexisting positive flux.  The dome therefore separates two topologically distinct magnetic domains.

The boundary between the two domains is easily identified in the EUV observations as a ``dark band'' that partially rings the newly emerged flux, indicated by the arrow in the 193\AA{} (upper left) image in \figref{fig:channelcomp}.  As \figref{fig:channelcomp} shows, the band is most easily seen along the southern portion of the active region core, and appears in all EUV wavelengths.  EUV loops to one side of the band are short, bright, and connect to the newly emerged positive flux, while those on the other are long, diffuse, and appear to connect in the large area of unipolar positive flux across the diffuse PIL.  We call the set of short bright loops the \emph{kernel}.

The properties of the band that separate the two domains are discussed in detail in \inlinecite{Scott:2013}.  Notable among these are continuous, persistent (lasting several days) blueshifts of $\sim 25\unit{km \ s^{-1}}$ observed in \emph{Hinode}/EIS.  As shown in \figref{fig:topo}, the band itself is approximately cospatial with the intersection of the coronal null's separatrix surface with the photosphere, \textit{i.e.} the ring of spine field lines to which one foot of each of the coronal null's separators attach (red lines in that figure).  In the MCT, this ring is a circuit of spine lines passing through both old and new negative flux concentrations.

\begin{figure}[ht]
  \begin{center}
    \includegraphics[width=\textwidth]{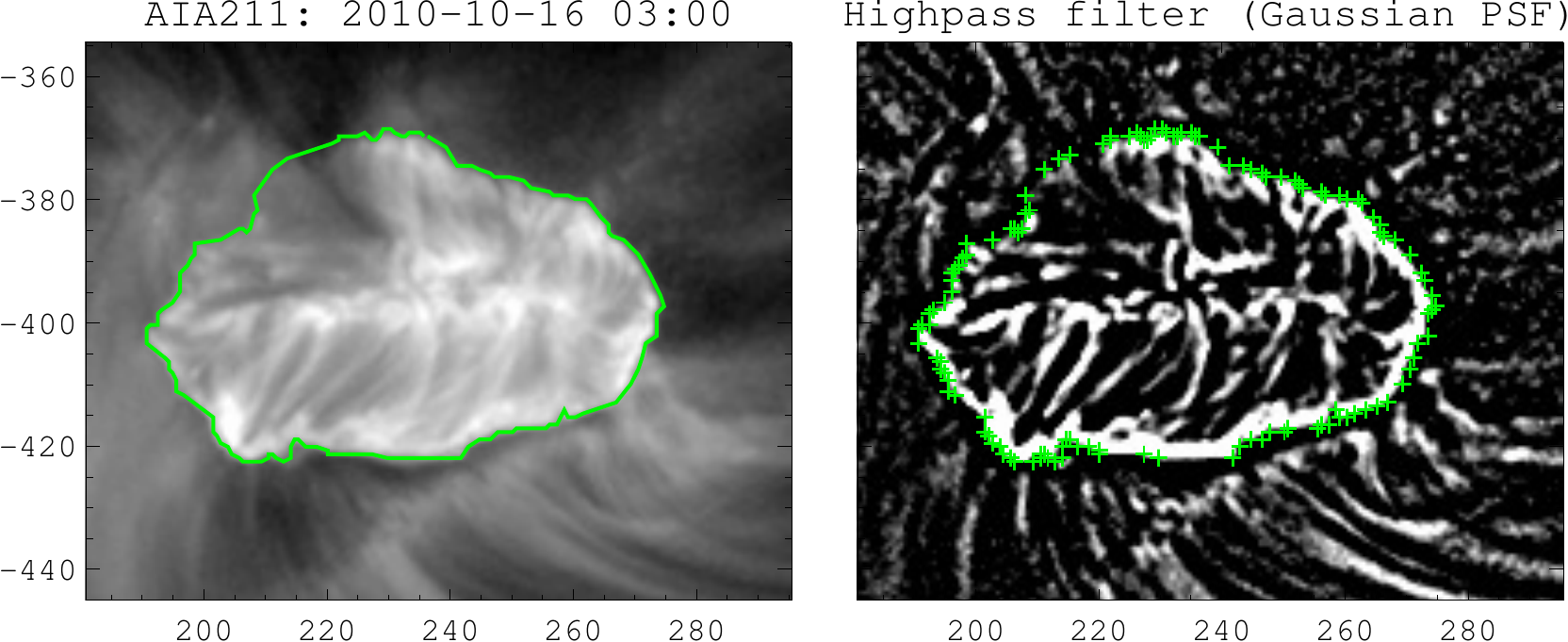}
    \caption[AR11112 EUV boundary]{Boundary of kernel flux and external flux, determined from EUV images.  Right: 211\AA{} data after convolution with a Gaussian highpass filter.  Green plusses show the by-eye chosen pixels that define the boundary.  Left: log-scaled 211\AA{} image with flux domain boundary superposed as a solid green line.}
    \label{fig:euvbnd}
  \end{center}
\end{figure}

To find precise spatial boundaries of the kernel at each timestep, we use data from the AIA 211\AA{} channel at every hour beginning at 14 October 2010 00:00 and ending on 17 October 2010 11:00.  We trace the boundary by eye in a highpass filtered version of each EUV image, as shown in \figref{fig:euvbnd}.  The highpass filtered image is created by convolving the image with a ten pixel by ten pixel Gaussian kernel with a standard deviation of 5 pixels and subtracting that from the original image.  After this process, the dark band itself is easily visible along the east, south, and west sides of the active region kernel.  The northern boundary is usually obscured by both the loops within the kernel and loops connected to the patch of highly concentrated preexisting negative flux just north of the emergence zone, labeled N1.  In the animation \figref{fig:channelcomp}, we see that the diffuse set of loops connected to N1 pass across the northern portion of the AR kernel like the beam from a lighthouse.

\figref{fig:euvbnd} shows the AIA 211\AA{} channel from \figref{fig:channelcomp}.  Displayed is the 16 October 2010 03:00 image in a logarithmic greyscale on the left and after convolution with the highpass filter on the right.  The selected points are shown as green plus marks in the right-hand image.  We select points on the inside of the dark band, at the edge of loops connecting to P1.  The boundary defined by these points is shown as a solid green line in the left-hand image.  A similar boundary is created for each timestep in our data series, as seen in the associated animation.

We can first identify the boundary at 14 October 2010 22:00.  This is approximately 17 hours after the first signs of emergence are visible in the magnetic field data, and 13 hours after the bright loops first become visible in the AIA:211\AA{}, XRT:Ti-poly, and XRT:Al-mesh data.

\section{\label{sec:rxrate}Inferred Reconnection Rate}
We overlay the boundary between the two coronal domains (loops connected to emerged positive flux, and loops connecting elsewhere) determined in \S\ref{sec:euv} over the closest-in-time radial magnetic field at each timestep.  See, for example, the thick line in \figref{fig:magevo}.  The intersection of this boundary with the contour of pre-emergence negative flux to the west of emergence (N2) provides us with a lower-bound estimate of the amount of reconnection between the old and newly emerged flux.  This is quite a conservative estimate because we exclude any reconnection to N13 from our analysis, owing to that region's proximity to the emergence zone.  Furthermore, after Oct 15th at 23:00 we begin to see the EUV boundary pass through the contour for N1, north of the emergence zone, indicating additional reconnection with that preexisting concentration.  We will return to these considerations in more detail below.

\begin{figure}[ht]
  \begin{center}
    \includegraphics[width=\textwidth]{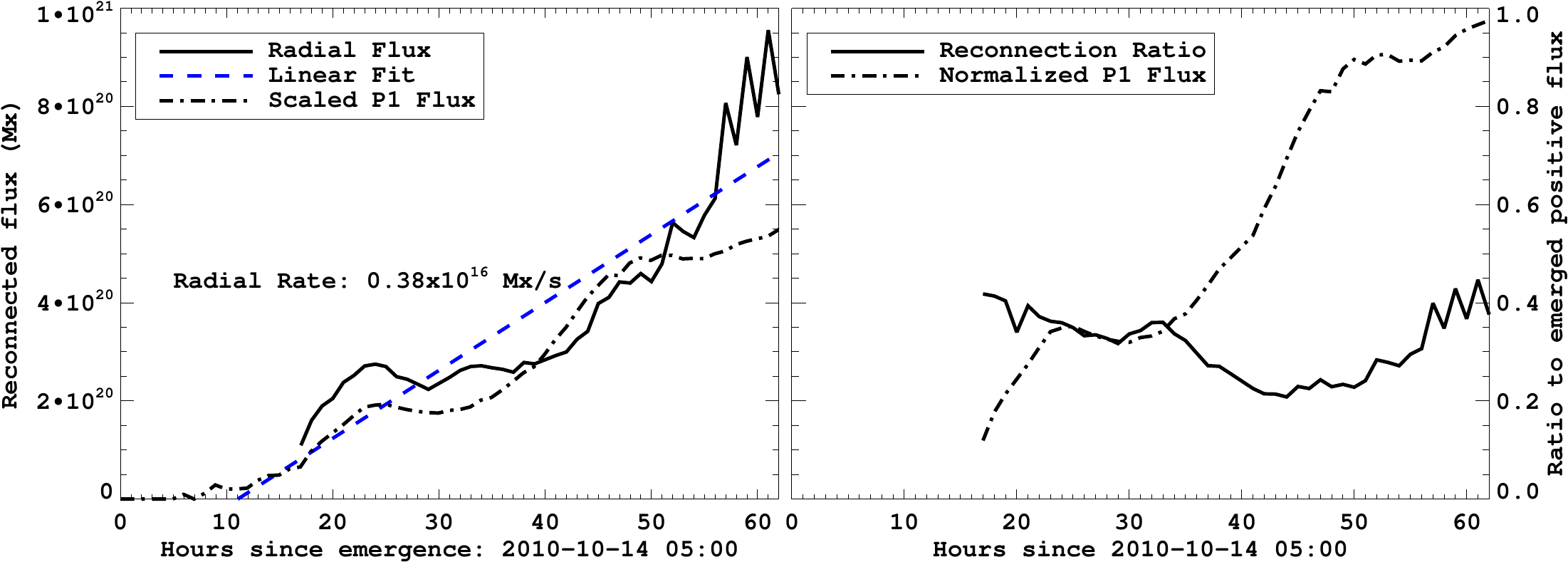}
    \caption[AR11112 inferred reconnection rate]{
Left: Estimated reconnected flux determined by the overlap between EUV boundary and the boundary of N2 from radial magnetic flux (solid black line) and a linear fit (dashed blue).  The total emerged positive flux is scaled by $0.25$ (dash--dot black).  Right: Ratio of the reconnected flux from the radial field to the total emerged positive flux (solid), and the total emerged positive flux (P1) normalized to its maximum value of $2.2\times10^{21}\unit{Mx}$ (dash--dot).}
    \label{fig:phi_rx}
  \end{center}
\end{figure}
\figref{fig:phi_rx} plots the amount of flux in the intersection of the EUV boundary and N2 boundary at each timestep.  Because the EUV boundary delineates field lines connecting to the newly emerged positive and those connecting to preexisting positive flux, and because N2 is a region containing only preexisting negative flux, the amount of flux in the intersection is a measure of the amount of reconnected flux at a given time.  From the linear fit to the amount of reconnected flux (dashed blue line, left panel of \figref{fig:phi_rx}), we determine a relatively steady reconnection rate of $0.38\times 10^{16}\unit{Mx \ s^{-1}}$ over about 2 days.  This corresponds to a characteristic EMF of $\mathcal{E}=\frac{d\Phi}{dt}=38\times 10^6$ volts within the reconnection region.

We can estimate the fraction of emerged flux that has reconnected at a given time by taking the ratio of the reconnected flux in N2 to the total amount of newly emerged positive flux P1.  This ratio is shown as the solid line in the right side plot of \figref{fig:phi_rx}, and varies between $25\%$ and $46\%$ of the emerged flux.  The amount of total emergence is depicted as a dashed--dot line, which is the flux of P1 normalized to its maximum value of $2.2\times10^{21}\unit{Mx}$.  The absolute amount of emergence is not important for this comparison so much as the relative rate of emergence at different times.  The ratio may vary in phase with the rate of emergence, and therefore with the rate at which the free energy density increases in the coronal field, though this is difficult to establish precisely.  

Consider the decrease in the ratio from $t=30$ to $t=50$ hours.  During this time, the fraction of reconnected flux drops by less than half, while the amount of emerged flux more than doubles.  We see that the reconnection rate may be tracking the emergence rate to some degree, but that coupling is not simple.  Indeed, from the linear fit of the reconnection rate in the left plot of \figref{fig:phi_rx}, it may take at least 11 hours after flux emergence before any existing field reconnects with new field.  Delay between emergence and reconnection has been reported in previous observations \cite{Longcope:2005,Zuccarello:2008}, and may be a general feature of flux emergence.

Secondly, some of the variations we see in the amount of emerged flux are consistent with the diurnal variations in the measured magnetic field of HMI data \cite{Liu:2012}.  See in particular the ``humps'' in the dashed--dot line of normalized P1 emergence in \figref{fig:phi_rx}, around $t=25$ and 50 hours.  Because we are primarily concerned with the average reconnection rate over time, and because these variations equally affect the measured magnetic field throughout the FOV, we will not concern ourselves with this added source of uncertainty.

Finally, around $t=42$ hours after emergence, loops begin unequivocally to form between P1 and N1 (see the animation of \figref{fig:channelcomp}).  These are excluded from our analysis for the reasons discussed when defining the EUV boundary in the previous section.  Reconnection may already have occurred by $t=42$ between the two regions, although it is difficult to tell.  Parts of N1 form the footpoints of loops that appear to arch over the top of the emerging flux to connect with positive flux across the large scale PIL, obscuring the interaction between N1 and P1 until $t=42$.  For this reason we provide the conservative estimate limited to reconnection with N2.  

Regardless of these ambiguities, the total amount of reconnection between P1 and N1 is small, about $11\%$ of that between P1 and N2.  By the end of the time range we consider, $62$ hours after first emergence, the intersection between the EUV boundary and N1 contains $1.0\times 10^{20}\unit{Mx}$, compared to $8.9\times10^{20}\unit{Mx}$ for the intersection with N2.  The intersection of the EUV boundary with N13 contains $1.7\times10^{20}\unit{Mx}$.

The amount of emerged flux shown as the dot--dashed line in the right hand plot of \figref{fig:phi_rx} begins around $t=5$ hours after first emergence.  This start time is simply due to the thresholds we set in the creation of our mask structures.  It is unlikely to affect our linear fit to the reconnection rate, which shows first reconnection at $t=11$, because we are unable to measure any reconnected flux until $t=17$, 9 hours after P1 rises above the mask thresholds, and 17 hours after first noted emergence in the vector field data.

\begin{figure}[ht]
  \begin{center}
    \includegraphics[width=\textwidth]{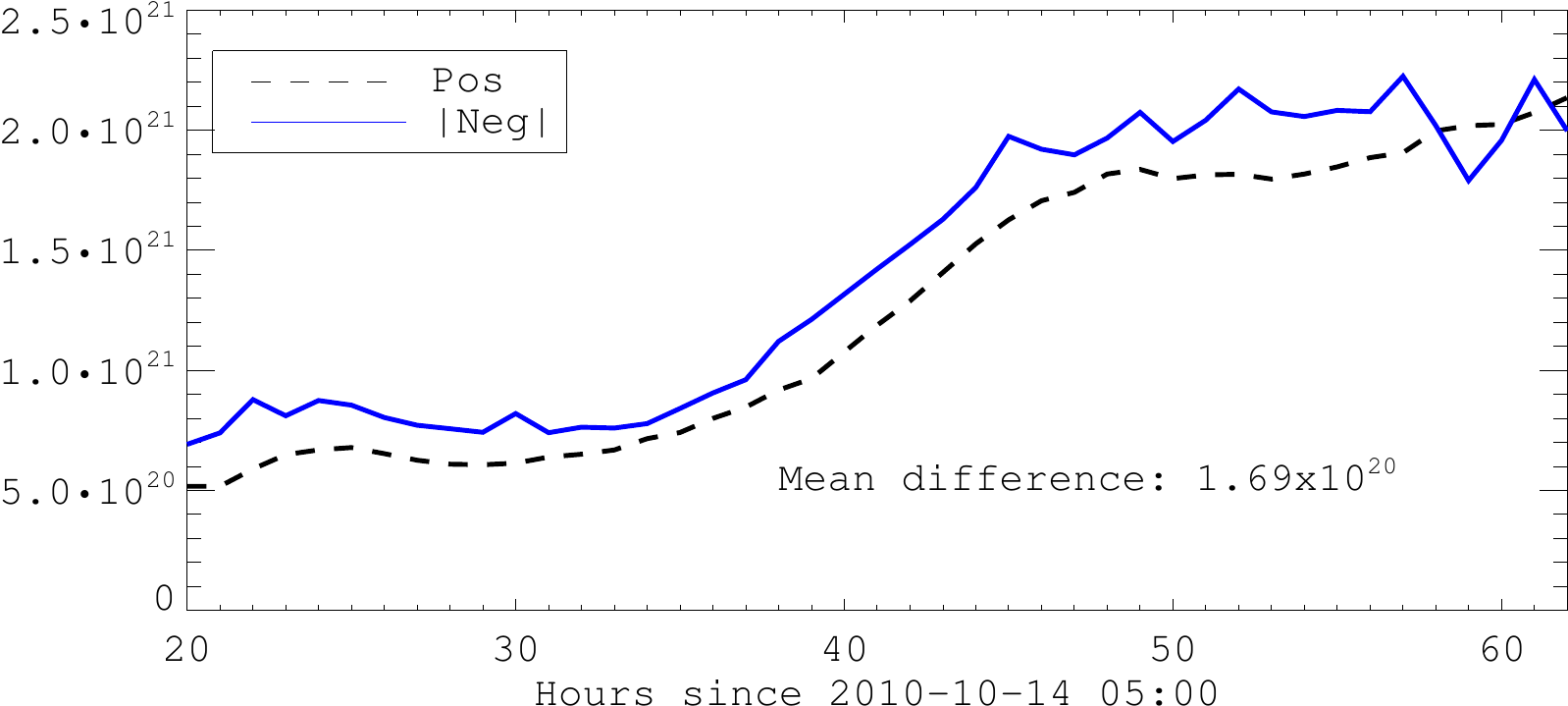}
    \caption[Flux balance within EUV boundary for AR11112]{Total positive (black dashed) and absolute value of negative (solid blue) flux within the EUV boundary.  All positive flux within the boundary is due to emergence, while the negative flux is a mix of preexisting and emerged flux.  The mean difference between the two curves is $1.69\times10^{20}\unit{Mx}$ with a variance of $\approx 1.0\times10^{20}\unit{Mx}$, between $5-10\%$ of the total.}
    \label{fig:fluxbalance}
  \end{center}
\end{figure}

The total amount of signed flux within the EUV boundary is well balanced, as shown in \figref{fig:fluxbalance}.  This value is only determined for times when we can observe the boundary.  We find a mean signed flux within the boundary of $1.69\times10^{20}\unit{Mx}$, with a variance of $\approx 1.0\times10^{20}\unit{Mx}$.  The consistency of the imbalance over time as new flux emerges indicates that we have accurately captured the evolution of this system.  Because our boundary passes through negative flux regions consisting of both emerged and preexisting flux, and because reconnection necessarily transfers flux one-to-one between flux domains, this is a further indication that the feature we observe in EUV is in fact the separatrix surface of the coronal null; flux within such a surface must be balanced, even as the surface changes due to continued emergence and reconnection.

\section{\label{sec:xrt}Temperature, Emission Measure, and Radiated Power}
We generate maps of temperature $T$ and volumetric emission measure $EM$ at each pixel in the XRT FOV using the filter ratio method described in \inlinecite{Narukage:2011}.  This analysis assumes an isothermal plasma.  Ratios of different filters provide more or less tightly constrained temperatures, with some ratios being ill-defined (multivalued) in certain regions, or with EM highly dependent on the assumed isotope abundances.  The ratio of Ti-poly to Al-mesh filters provides one of the cleanest temperature and emission measure indicators for expected plasma parameters in nonflaring active regions.  As discussed in that paper, the filter ratio method for a broadband instrument like XRT should correspond to a mean plasma temperature, weighted by the DEM of the emitting coronal plasma.  For this reason, it should result in a reliable measure of the radiated power.

We use a modified version of the \fnc{xrt\_teem.pro} program available in SSW to derive the $T$ and $EM$ maps.  The modified version, described in \inlinecite{Takeda:2012}, allows the user to analyze XRT filter ratios with assumptions of plasmas with either coronal, photospheric, or hybrid elemental abundances.  The different choices of abundances affect the calculation of XRT temperature responses and the resulting filter ratios.  Because the present work focuses on actively emerging flux, we use the hybrid model, with abundances based on \inlinecite{Fludra:1999}.  The EM in each pixel is given by
\begin{equation}
  \label{eq:em}
  EM= \int n_e^2 dV\, ,
\end{equation}
where $n_e$ is electron density within the pixel and the integral is over the volume of emitting plasma $V$.  The total power radiated by this plasma is
\begin{equation}
  P = \int \Lambda(T)\, n_e^2\, dV = \Lambda(T)\, EM(T)\,  ,
\end{equation}
where the final step assumes the entire emitting volume is at a single temperature $T$ and $\Lambda(T) = a(T) T^{f(T)}$ is the radiative loss function of \inlinecite{Klimchuk:2001}, summarized in Table \ref{tab:radloss}.  We then calculate the observed radiated power $P$ by
\begin{equation}
  \label{eq:power}
  \log (P) = \log(EM) + \log (a) + f(T) \log(T)\,  .
\end{equation}

\begin{center} 
\begin{table}[h]
\caption{\label{tab:radloss}  Parameters of the radiative loss function $\Lambda(T)= a(T) \times T^{f(T)}$ given by \protect\opencite{Klimchuk:2001}.}
   \begin{tabular}{ccc}
     \hline\hline\\
     Temperature range (K) & $a(T)$ & f(T)\\
     \hline 
     $\phantom{T_0< }T  < T_{0} = 1.0000\times10^4 $ & $0.0000$      &$ 0            $  \\
     $T_0< T < T_1 = 9.3325\times10^4$ & $1.0909\times10^{-31}$      &$+ 2.          $  \\
     $T_1< T < T_2 = 4.67735\times10^5$   & $8.8669\times10^{-17}$   &$-1.           $  \\
     $T_2< T < T_3 = 1.51356\times10^6$   & $1.8957\times10^{-22}$   &$ 0            $  \\
     $T_3< T < T_4 = 3.54813\times10^6$   & $3.5300\times10^{-13}$   &$-\frac{3}{2}  $  \\
     $T_4< T < T_5 = 7.94328\times10^6$   & $3.4629\times10^{-25}$   &$+\frac{1}{3}  $  \\
     $T_5< T < T_6 = 4.28048\times10^7$   & $5.4883\times10^{-16}$   &$- 1.0         $  \\
     $\phantom{T_0< }T_6< T$\phantom{$_6 = 4.28048\times10^7$} & $1.9600\times10^{-27}$   &$+\frac{1}{2}  $  \\
     \hline
   \end{tabular} 
 \end{table}
\end{center} 

XRT suffers from a time-varying contamination that collects on the CCD, obscuring about $5\%$ of the area in the form of condensation-formed spots \cite{Narukage:2011}.  While the locations of these spots are known, their wavelength-dependent opacities and time dependent thicknesses are not.  In order to compensate for this in our analysis, we smoothly interpolate the radiated power along the boundary of each spot into the spot's interior.  To do so we iteratively call the standard IDL \fnc{smooth} function on the power data, which convolves the data with a 2D boxcar function.   At each iteration we apply a 3 pixel by 3 pixel boxcar kernel, and we only update the contaminated pixels so that no uncontaminated data are affected.  The smoothing converges by 500 iterations, at which point the value of each pixel within a contaminated spot is the average of the value at each pixel on the spot's boundary, weighted by the distance to each boundary pixel.  This provides a more reasonable estimate of the radiated power than simply setting the value within the spots to zero.  

We estimate the error in the radiated power in each pixel as
\begin{equation}
  \Delta P = \frac{P}{EM}\Delta EM\,  ,
\end{equation}
where the error in the temperature and emission measure is calculated as in \inlinecite{Narukage:2011} due a photon noise threshold of $50\%$.  Because the radiative loss function is an emperically defined quantity and we expect the error in emission measure to dominate anyway, we take $\Delta \Lambda(T)$ to be 0.

\begin{figure}[ht]
  \begin{center}
    \includegraphics[width=\textwidth]{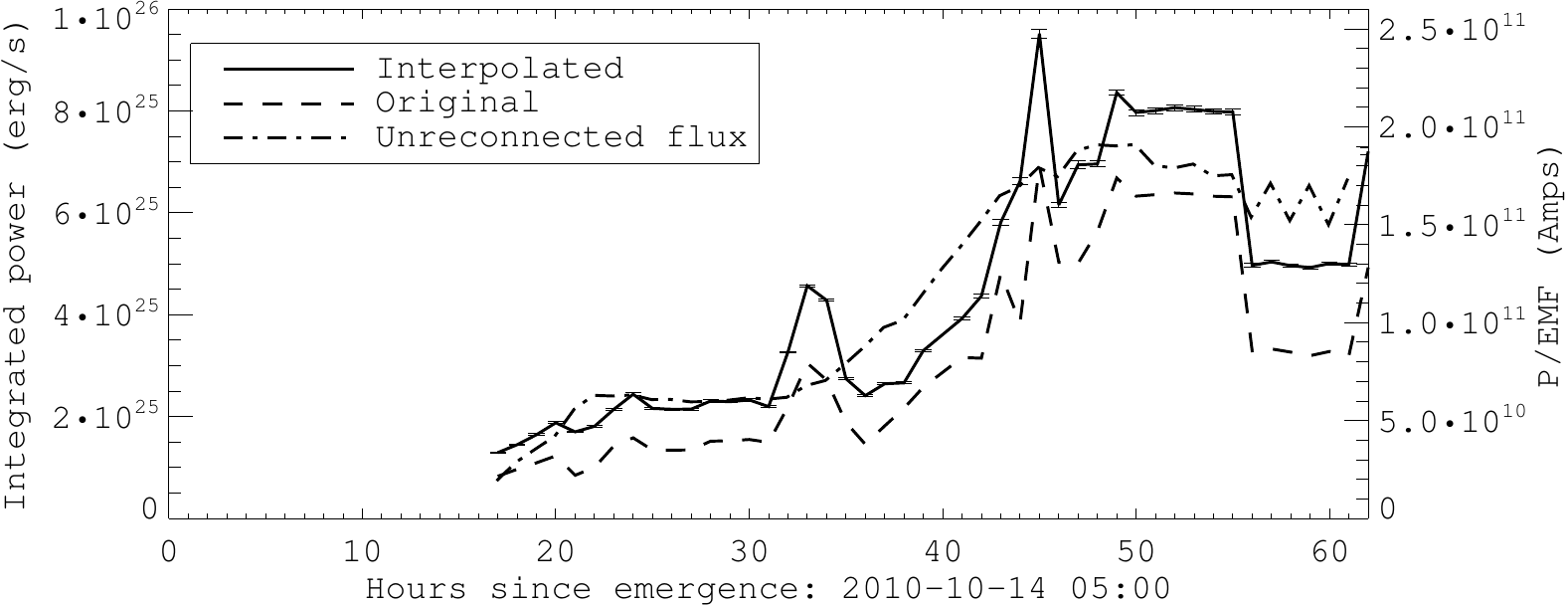}
    \caption[AR11112 inferred radiated power]{Left axis: Radiated power derived from XRT filter ratios, from Eq.~\ref{eq:power}, in $\unit{erg\ s^{-1}}$.  The solid line uses data interpolated into contaminated pixels, and the dashed line uses the original data.  Right axis: characteristic current in the reconnection region assuming a constant $\mathcal{E}=0.38\times 10^{16}\unit{Mx \ s^{-1}} = 38\times 10^{6}\unit{V}$.  The dash--dot line shows the unreconnected emerged flux, derived as the difference between flux in P1 and the reconnected flux of \figref{fig:phi_rx}, and then scaled by a factor $x=0.487\times10^{5}\unit{erg\ s^{-1} Mx^{-1}}$.  See text for a discussion of the drop near t=55.}
    \label{fig:power}
  \end{center}
\end{figure}

\figref{fig:power} shows the total power radiated by the plasma from pixels within the same boundary defined using the EUV data.  The solid line is the total after the interpolation described above, while the dashed line is the total setting each contaminated pixel to zero.  Interpolation increases the estimated radiated power by a roughly constant factor of $50\%$, consistent with the ratio of contaminated to uncontaminated pixels within the EUV defined boundary.  The power rises from $2$ to $8\times10^{25}\unit{erg \ s^{-1}}$ as the active region emerges.  The summed error for all XRT pixels in the EUV defined boundary (the number of pixels increases from $\sim 1200$ to $3500$ over the timeseries) is displayed as vertical bars for the interpolated power (solid line) discussed above.  The error from photon noise varies between $1.3-9.5\times 10^{23}\unit{erg \ s^{-1}}$, about $1\%$ of the power at each time, but is completely dominated by the more systematic $\sim 50\%$ increase due to the contaminated pixels.

Dividing the power by the mean reconnection rate, $\dot{\Phi}=0.38\times 10^{16}\unit{Mx \ s^{-1}}$, converts the radiated power to units of erg/Mx, equivalent to 10 Joules/Wb or 10 amps. The right axis shows the value of the power scaled this way.  This would be the current needed to produce the observed power as electromagnetic work with the mean electromotive force.  Reconnection flux transfer, $\dot{\Phi}$, across a current sheet of net current $I$, will {\em release} energy at the rate $I\dot{\Phi}$.  Since only a fraction of the energy released by reconnection will be promptly radiated, the values on the right axis of \figref{fig:power} could be taken as a lower bound on the current in the sheet surrounding the newly emerged flux.  On the other hand, the flux transfer rate in the denominator is itself a lower bound, so the current lower bound may be somewhat liberal. 

Interpreted as described above, the increasing curves of \figref{fig:power} suggest an increasing current in the sheet surrounding the emerging flux.  This increase occurs in spite of the continual reconnection which is observed to occur.  We must therefore conclude that the amount of {\em unreconnected flux} (responsible for the sheet's existence) is increasing as well.  This conclusion is consistent with the finding above that only 20\,--40\,\% of the emerged flux has been reconnected (see right plot of \figref{fig:phi_rx}).

The dash--dot curve in \figref{fig:power} shows the unreconnected flux found from the difference between the total flux in $P1$ and the reconnected flux shown as a solid curve in \figref{fig:phi_rx} left.  This unreconnected flux is scaled by a factor, $x=\dot{\Phi}/8\pi\ell$, to plot against power.   We find the best match to the corrected power (solid curve) using $\ell=31\unit{Mm}$.  This relatively good fit suggest that the net sheet current scales with unreconnected flux with a factor $\ell$.  This current-flux scaling factor is, by definition, the self-inductance (up to some dimensionless factor).  It is noteworthy that it is not grossly different than the linear dimension of the dome along which the current sheet presumably flows.

Note that there was an 11.5 hour gap in the XRT data between $t=50$ and $61.5 \unit{hrs}$ as XRT observed a different area of the Sun.  This creates the two flat portions of the calculated power for timesteps close to those two data points.  Additionally, at $t=61.5$ the kernel was obscured by a particularly large patch of contaminated CCD pixels, which likely led to the significantly decreased radiated power determined at those times.

The total energy radiated away by the system over the course of our analysis is determined by summing the radiated power over time.  Using the interpolated XRT data (solid curve in \figref{fig:power}), the system radiates $7.2\times 10^{30}\unit{ergs}$ over the 45 hours between establishment of the EUV boundary and the GOES M2.9 flare (summing only the uncontaminated pixels yields $5.5\times 10^{30}\unit{ergs}$).

\section{\label{sec:discussion}Discussion}

In this work we present the first (to our knowledge) quantitative measurement of the amount of transferred flux and the rate of reconnection during quiescent reconnection between an emerging flux bundle and surrounding, preexisting field.  The rate of reconnection is approximately steady at $0.38\times 10^{16}\unit{Mx \ s^{-1}}$ over the course of two days, though as noted above this is an underestimate because we focus only on one region of unambiguously preexisting negative flux.  We observe EUV loops connecting to other negative regions as well, but often the EUV boundary is difficult to determine in these locations or there is some mix between emerged and preexisting negative flux. 

While AR11112 does produce an M3.0 flare on Oct 16th, 2010, the implied reconnection we focus on here predates the flare by several days.  It produces little if any observable flaring activity of its own, such as increases in the GOES light curve or chromospheric flare ribbons.  Further, we do not observe any sudden jumps in the amount of reconnected flux.  Based on these observations, we conclude that this is indeed an instance of quiescent reconnection.

Our analysis is conceptually similar to those attempting to determine the amount of flux involved in two ribbon flares.  \inlinecite{Qiu:2010} perform a similar measurement by summing the flux for magnetogram pixels cospatial with UV flare ribbons during the well studied Bastille Day (2000) two ribbon flare.  Their event lasts for less than an hour and their measured rate varies by an order of magnitude, between $10^{18}$ and $10^{19}\unit{Mx \ s^{-1}}$ during this time, resulting in about $10^{22}\unit{Mx}$ total transferred flux.  \inlinecite{Kazachenko:2010} apply a similar method to another heavily studied flare, the Halloween flare of 2003, again finding on order $10^{22}\unit{Mx}$ of flux transferred during the flare.  These two events are among the strongest flares ever recorded for the Sun \cite{Schrijver:2012} -- GOES classes X5.7 and X17, respectively -- and several orders of magnitude larger in GOES class than any event in AR11112 during the time of our analysis.  Direct comparison of reconnection rates and total flux transfers between these cases is thus problematic.

\inlinecite{Tarr:2013} used the conceptually unrelated free energy minimization scheme to estimate flux transferred during the M6.6, M2.2 and X2.2 flares of February 2011 in AR11158.  They found the flux involved in each flare to be $4.2\times 10^{20}\unit{Mx},\ 2.0 \times 10^{20}\unit{Mx} , \hbox{and } 21.0 \times 10^{20}\unit{Mx}$, respectively, 1\.--\,2 orders of magnitude less than that for the Bastille Day and Halloween flares.  The reconnected flux we measure in AR11112 of $\approx 1\times10^{21}\unit{Mx}$ may be consistent with the amount of flux involved in a smaller GOES X-class flare.  At the same time, the total amount radiated energy originating within the EUV boundary over the time series is $7.2\times10^{30}\unit{ergs}$, also consistent with a large M- or small X-class flare \cite{Longcope:2010,Kazachenko:2012}.  In terms of the reconnection and radiative processes ongoing throughout AR11112's emergence, this entire event \emph{is} an X-class flare, but simply takes 45 hours instead of the 30 minutes observed, for instance, in \inlinecite{Qiu:2009}.

Our conclusion raises a basic question.  Why does reconnection occur very rapidly in some cases, namely flares, yet far more slowly in other cases, like the present?  Our analysis suggests that the difference does not lie in the net flux or net energy involved, since these are similar for our quiescent case and in some sizable flares.  Applying the energy {\em vs}.~flux scaling from \S\ref{sec:xrt} to the flares would naturally lead to net currents comparable in both cases.  Thus the net current in the current sheet cannot be the factor discriminating between fast and slow magnetic reconnection.  The difference may instead lie in the geometry of the current sheet, or some aspect of its dynamics.  This fundamental question concerning magnetic reconnection will need to be addressed in future work.

Another puzzling result is the apparent delay between the time of photospheric emergence and any measureable amount of reconnected flux, both during the initial emergence and subsequent surges.  Recall that we see our first signatures of flux emergence in the photosphere around 14 October 2010 05:00 as rapidly expanding regions of highly inclined magnetic field.  4 hours later we detect our first signatures in the 211\,\AA{} images as locally enhanced emission measure.  Three hours after that we see the first well defined EUV loops.  Finally, at 22:00, 17 hours after emergence, we observe a fully formed kernel of bright loops bounded by a persistent dark band.  By taking a linear fit to the next 45 hours of our measured reconnected flux, we infer that reconnection began 11 hours after first emergence.  

\inlinecite{Longcope:2005b} used a rather different method (yet also using EUV loops) to estimate a reconnection rate between two active regions, one emerging one preexisting.  They find evidence of reconnection between 6 and 24 hours after emergence.  In yet another study, \inlinecite{Zuccarello:2008} examined reconnection between an arch filament system that emerged along the PIL in a well established, though not yet diffuse, active region.  They determine a time delay of about 10 hours between emergence and the reconnection event, a small C-class flare.

Another important point is that flux constantly emerges through the photosphere, even in quiet sun regions during solar minimum.  Even this small scale emergence likely operates in a fashion similar to the larger, active region scale phenomena described in this study.  Much work has recently been devoted to the numerical study of emergence-induced X-ray jets and bright-points, and the related problem of network flux ``recycling'' times: the rate at which emerging small scale flux reconnects to form new footpoints for open magnetic field \cite{Schrijver:1997,Archontis:2008,Hagenaar:2008,Cranmer:2010,Moreno:2013}.  

In a recent recycling time investigation, \inlinecite{Cranmer:2010} study the energetic consequences of flux emergence and reconnection in the magnetic carpet (first described by \opencite{Schrijver:1997}).  They compare the recycling time for flux emergence (time required for flux to emerge from below the photosphere) and the rate at which closed flux becomes open.  The two timescales compare favorably (see Figure 10(a) of \inlinecite{Cranmer:2010}), and for regions of highly imbalanced initial flux, such as we have for AR11112, that timescale is about 11 hours.  This comparison should not be taken too far, as we have not shown the exact relation between their simulations and our observations.  It is, however, a comparison that should be considered further, as it is in general agreement with our findings in this work, as well as the other studies mentioned above which find a 10 to 12 hour delay between emergence and first observations of inferred reconnection.

Direct comparison between the present work and the most similar numeric simulations, those of coronal jets and bright points as in \inlinecite{Moreno:2013}, is difficult.  As is typical in this type of study, \inlinecite{Moreno:2013} focus on relatively small amounts of flux emergence ($\sim 10^{19}\unit{Mx}$) and shorter timescales (10s to 100 minutes), consistent with observations of X-ray jets.  However, they do see a slight delay between a substantial amount of emergence into coronal volumes and the change of field line connectivities between emerging and surrounding open field.  As they mention at the end of their \S7, it is difficult to derive a global rate of reconnection from their simulations.  Consider the connectivity of field lines passing through a given height above the emergence zone.  One gets a sense of such connectivity's evolution from their Figures 14 and 15.  We see this type of behavior in our observations as well, though this is not what we mean by reconnection, where two coronal flux domains swap footpoints, but is instead largely due to the expansion of new flux into coronal volume.  

\inlinecite{Pontin:2013} consider the related case of reconnection across a magnetic dome when there is no flux emergence.  The reconnection is initiated first by generating a field--aligned current in the vicinity of the coronal null in an analytic model, and second by advecting the parasitic polarity underneath the dome.  The evolution of their field line connectivites are broadly consistent with the observations we have presented in the present work.  A more detailed comparison between our observations and related numerical simulations must be given in a future work.

\acknowledgements
This work is supported by NASA under contract SP02H3901R from Lockheed--Martin to MSU.  We would like to acknowledge our use of NASA's Astrophysics Data System.


\end{article}
\end{document}